\documentclass[prl,showpacs,reprint,aps]{revtex4-1}
\usepackage{amssymb}
\usepackage{amsmath}
\usepackage[dvips]{graphicx}
\usepackage[usenames,dvipsnames]{color}
\usepackage[english]{babel}
\usepackage[latin1]{inputenc}
\usepackage{latexsym}
\usepackage{natbib}
\newcommand{\op}[1]{\hat{\mathbf{#1}}}
\newcommand{\oop}[1]{\tilde{\mathbf{#1}}}
\newcommand{\Tr}{\mathrm{Tr}}
\begin{document}
\title{Electronic entanglement in late transition metal oxides}
\author{Patrik Thunstr{\"o}m}
\author{Igor Di Marco}
\author{Olle Eriksson}
\affiliation{Department of Physics and Astronomy, Uppsala University, Box 516, SE-75120, Uppsala, Sweden}
\date{\today }
\begin{abstract}
Here we present a study of the entanglement in the electronic structure of the late transition metal monoxides - MnO, FeO, CoO, and NiO - obtained by means of density-functional theory in the local density approximation combined with dynamical mean-field theory (LDA+DMFT). The impurity problem is solved through Exact Diagonalization (ED), which grants full access to the thermally mixed many-body ground state density operator. The quality of the electronic structure is affirmed through a direct comparison between the calculated electronic excitation spectrum and photoemission experiments. Our treatment allows for a quantitative investigation of the entanglement in the electronic structure. Two main sources of entanglement are explicitly resolved through the use of a fidelity based geometrical entanglement measure, and additional information is gained from a complementary entropic entanglement measure. We show that the interplay of crystal field effects and Coulomb interaction causes the entanglement in CoO to take a particularly intricate form.
\end{abstract}
\maketitle

Entanglement is a fundamental aspect of quantum mechanics, responsible for a large range of complex phenomena not present in a classical setting. The entanglement of distinguishable particles was historically seen as something spooky, but is now considered a valuable resource. It plays an essential role in quantum information theory, and has been studied in great detail both theoretically and experimentally~\cite{amico08rmp80:517}. Entanglement of indistinguishable particles has received less explicit attention, but it has been studied indirectly in both the quantum chemistry and condensed matter communities. Describing the electronic structure of materials with a pure separable state, represented as a single Slater determinant, is at the very heart of the modern computational approaches, and the inability to do so is known under the term `correlation'. The term encompasses both classical and quantum correlations (entanglement), where the former results in mixed states and the latter in entangled states. The presence of entanglement is usually considered as a computational complication as it prevents the use of these standard approaches. The development of reliable computational methods and entanglement measures are of key importance to turn also the entanglement of indistinguishable particles from a complication into a potential resource.

Well known examples of strongly correlated materials are the late transition metal monoxides (TMO) -- MnO, FeO, CoO and NiO -- which have been under intense experimental and theoretical attention for a long time~\cite{mott37pps49:72,*brandow77ap26:651,*zaanen85prl55:418,*imada98rmp70:1039,anisimov91prb44:943,*kobayashi08prb78:155112,*jiang10prb82:045108,*rodl09prb79:235114,*fujimori84prb29:5225,*eder08prb78:115111,*kutepov10prb82:045105,ren06prb74:195114,*kunes07prb75:165115,*kunes08nm7:198,*korotin08ejpb65:91,*karolak10jesrp181:11,*yin08prl100:066406}. Here we report on a theoretical description of the late TMOs using the Local Density Approximation plus Dynamical Mean Field Theory (LDA+DMFT)~\cite{georges96rmp68:13,*held07ap56:829} where the effective impurity model is solved by Exact Diagonalization (ED)~\cite{caffarel94prl72:1545}. The quality of the results is assessed through a direct comparison between the computed density of states (DOS) and photoemission experiments (XPS/BIS). We then take advantage of the direct access to the local many-body ground state of the impurity problem to analyse the entanglement in detail.

\begin{figure*}
\includegraphics[width=8cm]{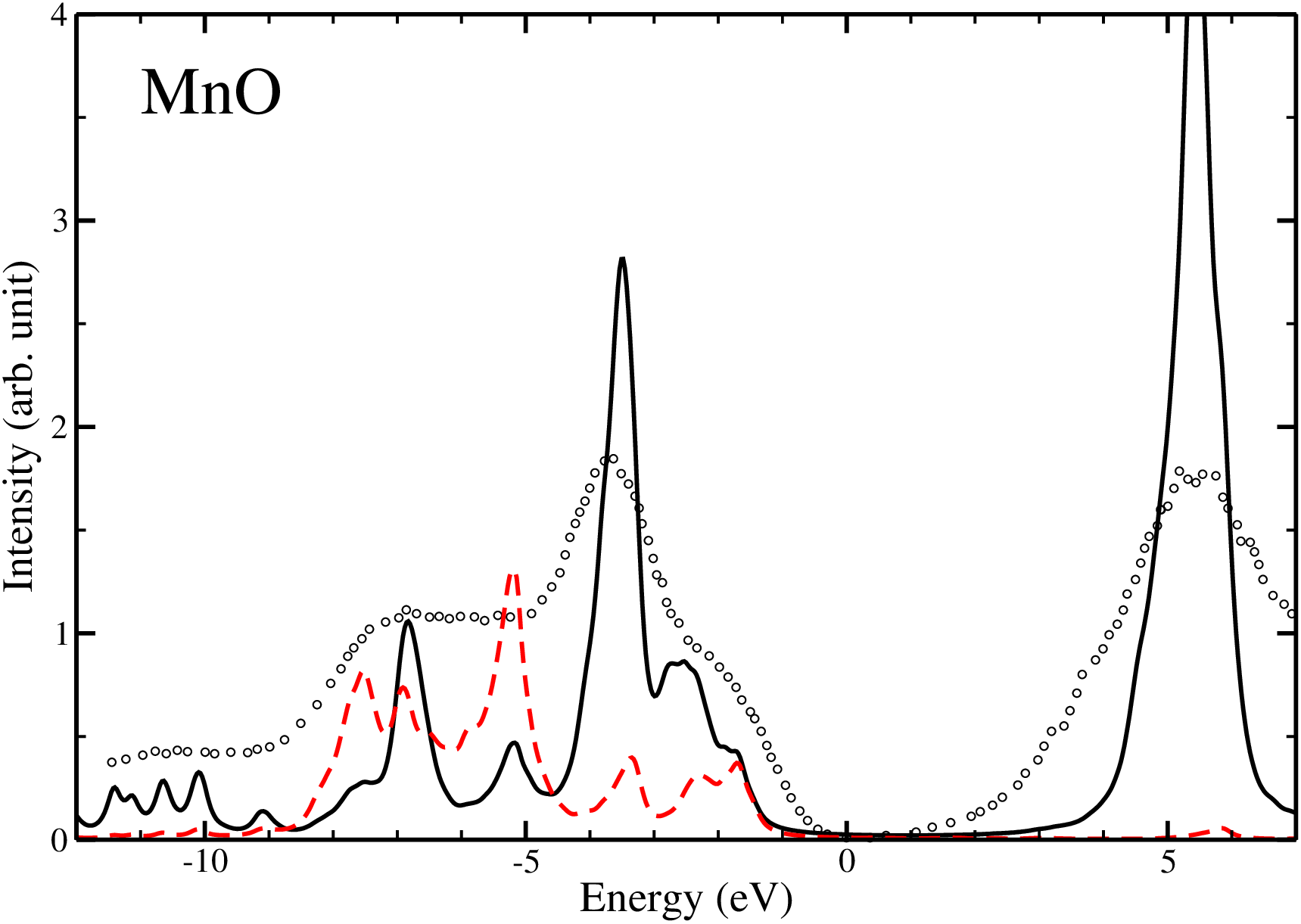}\hspace{1cm}\includegraphics[width=8cm]{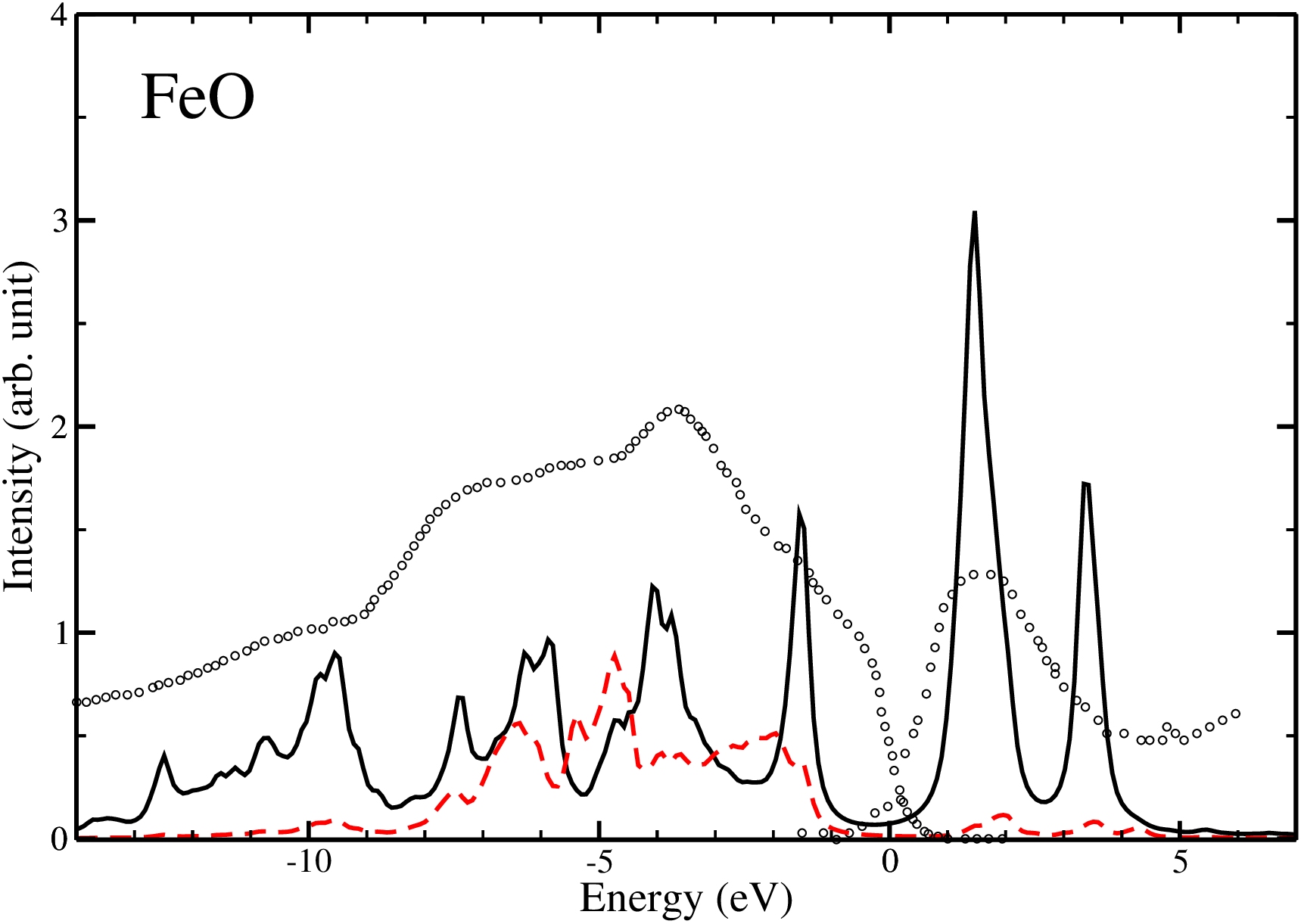}\\
\includegraphics[width=8cm]{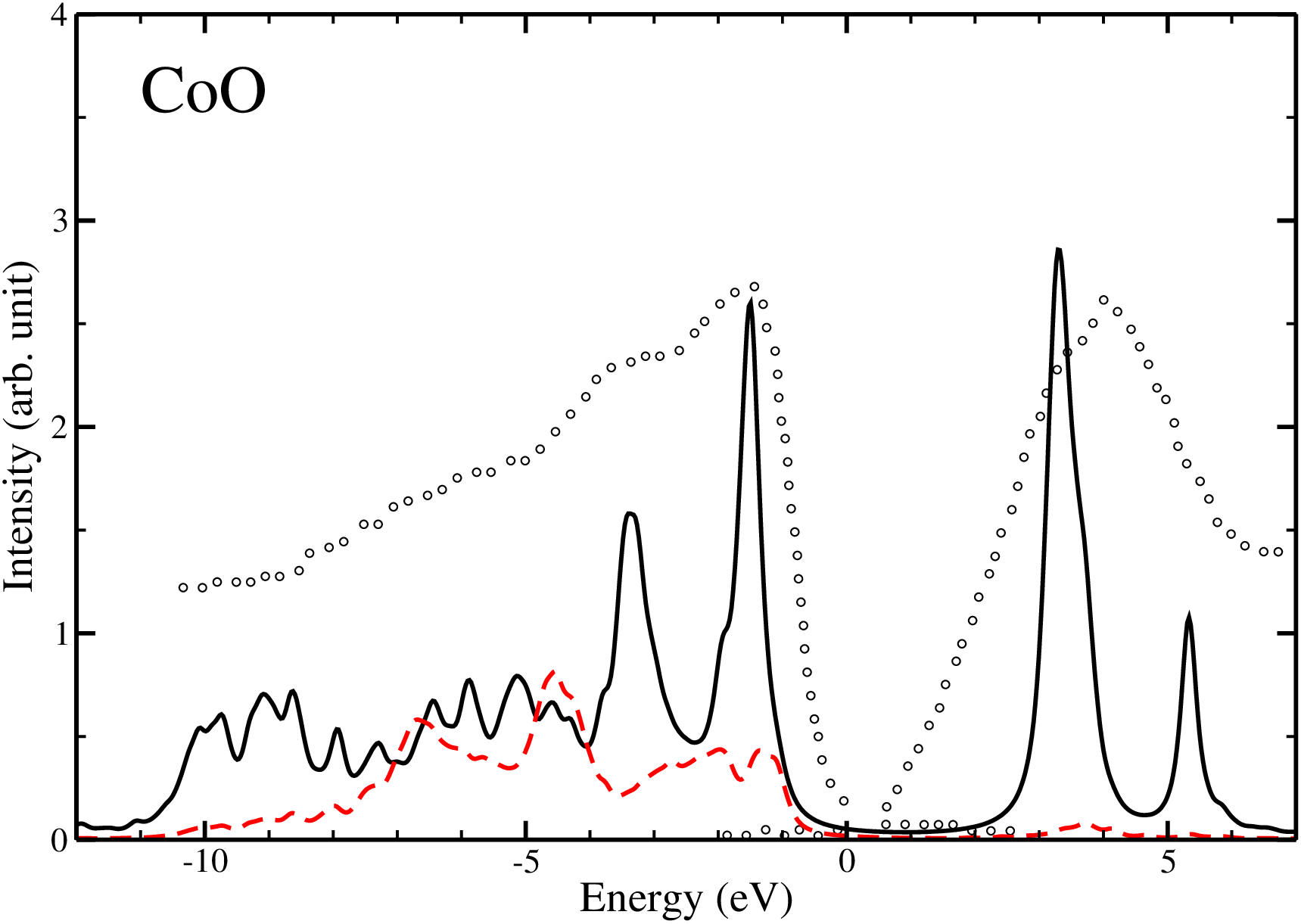}\hspace{1cm}\includegraphics[width=8cm]{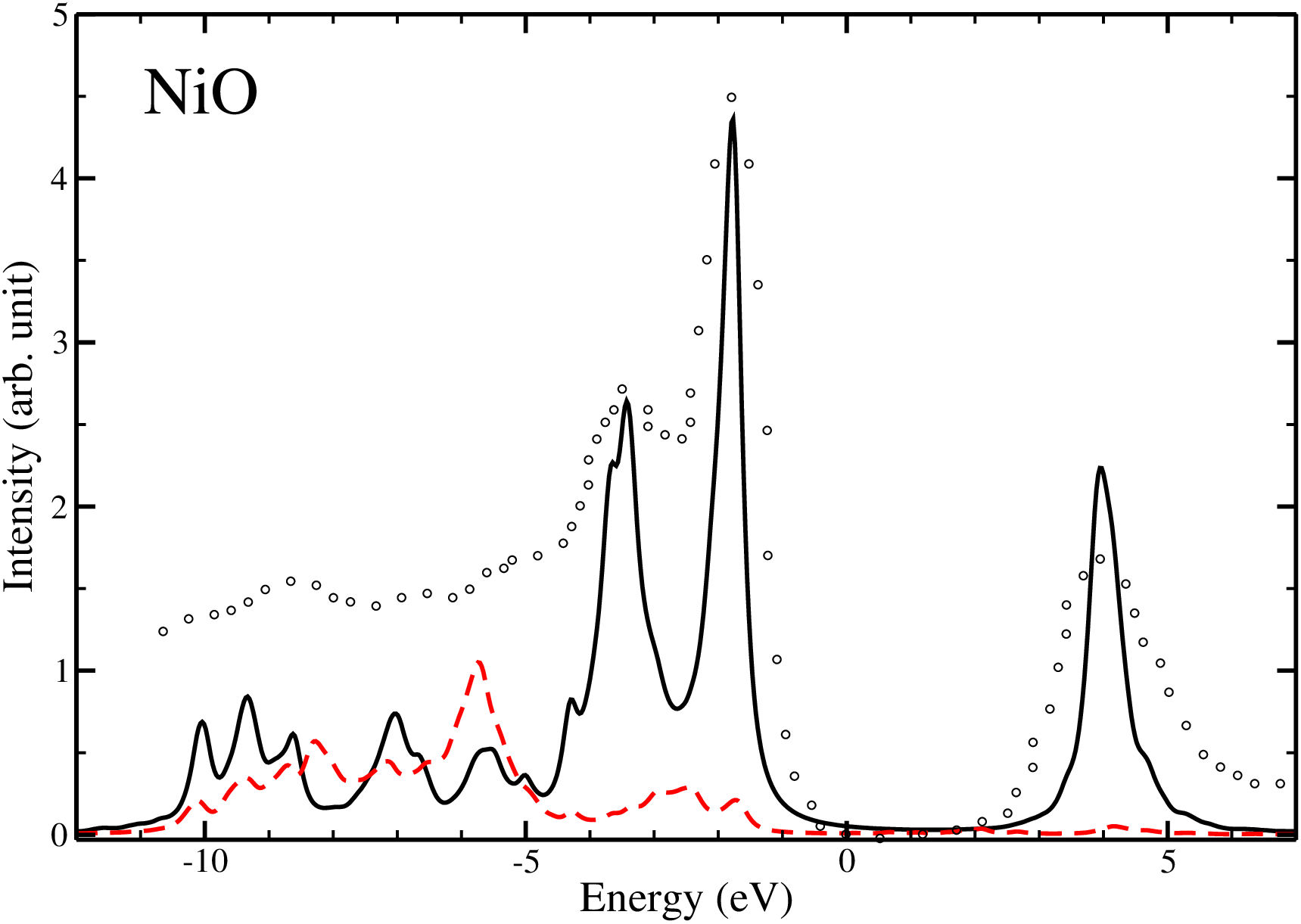}
\caption{(Colour on-line) Spectral function of the TM 3d states (thick black lines) and O 2p states (dashed red lines) in MnO, FeO, CoO, and NiO, and corresponding XPS/BIS data (black circles)~\cite{elp91prb44:1530,*prince05prb71:085102,*zimmermann99jpcm11:1657,*elp91prb44:6090,*sawatzky84prl53:2339}. The Fermi level is at zero energy.\label{fig:tmo}}
\end{figure*}

The LDA+DMFT scheme is built around the mapping of the local lattice problem to an effective impurity problem. The impurity system is described in ED through the local projected LDA Hamiltonian $\op{H}^{LDA}$, a double counting term $\op{H}^{DC}$, the on-site Coulomb interaction $\op{U}$, and a few auxiliary bath states, giving the Hamiltonian
\begin{eqnarray}
\op{H}^{ED} & = &\sum_{ij} \left( \oop{H}^{LDA}_{ij} - \oop{H}^{DC}_{ij} \right) \op{c}^\dagger_i\op{c}_j + \frac{1}{2}\sum_{ijkl} \oop{U}_{ijkl} \op{c}^\dagger_i\op{c}^\dagger_j\op{c}_l\op{c}_k \nonumber\\
 & + & \sum_{im}\left(\oop{V}_{im} \op{c}^\dagger_i\op{c}_m + \mathrm{H.c.} \right) + \sum_{m} \oop{E}_m \op{c}^\dagger_m\op{c}_m,\label{eqn:hed}
\end{eqnarray}
where the indices $i,j,k,l$ run over the local correlated orbitals and $m$ runs over the auxiliary bath states. The energies $E_m$ and the hybridization strength $V_{im}$ of the auxiliary bath states mimic the hybridization between the TM-3d and the O-2p and O-2s orbitals. The Hamiltonian of the finite system is diagonalized numerically to produce an analytical self-energy. The calculations were carried out in the paramagnetic (PM) phase, using a finite temperature ($\beta = 0.00173$ Ry) fully charge self-consistent LDA+DMFT implementation~\cite{savrasov04prb69:245101,thunstrom09prb79:165104,grechnev07prb76:035107,*dimarco09prb79:115111,*granas12cms55:295,RSPt_book}. Further technical details, including the parametrization of $\oop{U}_{ijkl}$ and the fitting of $E_m$ and $V_{im}$, can be found in the Supplemental Material~\cite{edsupplemental}.

In Fig. \ref{fig:tmo} the calculated projected spectral function of the TM-3d and O-2p states are compared to experimental XPS (electron removal) and BIS (electron addition) data~\cite{elp91prb44:1530,*prince05prb71:085102,*zimmermann99jpcm11:1657,*elp91prb44:6090,*sawatzky84prl53:2339} showing mainly the contribution of the TM-3d states, due to the photon energies used. The overall agreement for MnO, CoO, and NiO is excellent, and even minor experimental features like the high energy satellites are found in our theory. The CoO sample used in the experiment was doped with 1\% Li to avoid charging effects. However, this doping gives rise to Co$_3$O$_4$-like domains in the sample, which contribute to the early onset seen in the BIS spectrum
The projected spectral function of FeO show an acceptable agreement with the experimental spectrum, although the relative intensities could be improved. A greater concern is that the initial shoulder at -0.5 eV is missing and that the peak at 3.5 eV is not evident in the experimental data. However, this may be related to the fact that the experiment was performed with a non-stoichiometric sample (Fe$_{0.95}$O) including Fe$_3$O$_4$-like domains~\cite{bauer80mrb15:177,oku79jap50:6303}, in contrast to the conventional FeO unit cell used in the calculation. Nevertheless, it can not be ruled out that these features are beyond what can be described with the current method.

Given the satisfactory comparison of the spectral properties with experimental data, we now turn our attention to the entanglement in the thermal many-body ground state of the impurity problem, described by the density operator $ \op{\rho}^T = {e^{-\beta H^{ED}}} / {\Tr(e^{-\beta H^{ED}})}$. Before we start let us for clarity briefly introduce the concept of pure, mixed, separable, and entangled~\footnote{Here we refer to the entanglement between the electrons, and not to the concept of mode-entanglement~\cite{amico08rmp80:517}.} N-electron many-body states. A pure state is a state which can be described by a state vector $|\Psi\rangle$, while a mixed state requires the use of a density operator $\op{\rho}$. A mixed state is said to be classically correlated as its components are related through classical probabilities. A separable (non-entangled) pure state $|\Psi'\rangle$ can be written as a single Slater determinant~\cite{amico08rmp80:517}, while an entangled pure state $|\Psi\rangle$ requires a superposition of several Slater determinants. An entangled state is said to be quantum correlated as the superposition between its components is a quantum mechanical phenomenon. Finally, a separable mixed state can be written in a diagonal form $\op{\rho}' = \sum_i p_i |\Psi_i'\rangle\langle\Psi_i'|$ using only separable component states $|\Psi_i'\rangle$, while an entangled mixed state requires at least one entangled component~\cite{amico08rmp80:517}. A key point in the definition of mixed state entanglement is that a classical mixture of any two separable states remains separable.
 
There exist several ways to measure the entanglement in an N-electron many-body state~\cite{amico08rmp80:517}. In the case of a pure state $|\Psi\rangle$, we first look at the geometric entanglement measure~\cite{vedral97prl78:2275}
\begin{equation}
E_G[|\Psi\rangle\langle\Psi|] = 1 -  \max_{|\Psi'\rangle} \left| \langle\Psi'|\Psi\rangle \right|^2,\label{eqn:efpure}
\end{equation}
where $|\Psi'\rangle$ is restricted to be pure and separable. From a computational point of view this measure is natural since $|\Psi'\rangle$ corresponds to the best possible single Slater determinant description of the system. We perform the search for the optimal $|\Psi'\rangle$ by recursively removing one electron at the time from the natural orbitals of the many-body state~\cite{edsupplemental}. Although this procedure is in general not guaranteed to find the optimal $|\Psi'\rangle$, it is exact for separable states and for pure 2-electron systems, where it gives $E_G[|\Psi\rangle\langle\Psi|] = 1 - p_{max}$ with $p_{max}$ the largest eigenvalue of the corresponding one-particle reduced density matrix $\oop{\rho}_{ij} = \langle\Psi| c^\dagger_j c_i |\Psi\rangle$. 

In order to analyse the thermal density operator $\op{\rho}^T$ it is necessary to generalize the entanglement measure $E_G$ in Eq. (\ref{eqn:efpure}) to mixed states. This can be achieved by replacing the overlap with the fidelity~\cite{jozsa94jmo41:2315} between $\op{\rho}^T$ and any separable state $\op{\rho}'$,
\begin{equation}
E_G[\op{\rho}^T] = 1 - \max_{\op{\rho}'}\Tr\left[\sqrt{\sqrt{\op{\rho}^T}\op{\rho}'\sqrt{\op{\rho}^T}}\right]^2,\label{eqn:ef}
\end{equation}
where $\op{\rho}' = \sum_i p_i |\Psi_i'\rangle\langle\Psi_i'|$, $\sum_i p_i = 1$, and $|\Psi_i'\rangle$ are separable states. Performing the restrained maximization in Eq. (\ref{eqn:ef}) is in general a formidable task, as one has to consider the effect of mixing several non-orthogonal separable states. However, the problem can be simplified by noting that $\op{H}^{ED}$ of a PM system with negligible spin-orbit interaction commutes with $\op{S}^2$, $\op{S}_z$ and its ladder operators $\op{S}^\pm$. This implies that there is a common eigenbasis in which the many-body eigenstates $|\Psi^s_{i,m_s}\rangle$ can be indexed by \mbox{$s(s+1) = \langle\op{S}^2\rangle$}, and \mbox{$m_s = \langle\op{S}_z\rangle$}, and that the eigenvalues $E^s_i$ are independent of $m_s$. If the system has a well-defined spin moment, e.g. due to a strong on-site Hund's coupling, its thermal density matrix is composed of a mixture of several degenerate eigenstates, all with the same spin quantum number $s$
\begin{equation}
\op{\rho}^T = \sum_i p_i \op{\rho}^T_i = \sum_i \frac{p_i}{2s+1} \sum_{m_s = -s}^{s} |\Psi^{s}_{i,m_s}\rangle\langle \Psi^{s}_{i,m_s} |.\label{eqn:rhot}
\end{equation}
We also observe that the spin-coherent operator $\op{S}_{\lambda} = \exp(\lambda \op{S}^-) (1 + |\lambda|^2)^{-\op{S}_z}$ conserves the entanglement~\cite{edsupplemental} of the maximally spin-polarized state $|\Psi^s_{i,s}\rangle$. Setting $\lambda_m = \exp(2i\pi m/(2s+1))$ yields
\begin{equation}
\op{\rho}^T_i = \sum_{m = -s}^{s}\frac{\op{\rho}^{\lambda_m}_i}{2s+1} \equiv \sum_{m = -s}^{s} \frac{\op{N}_z \op{S}_{\lambda_m}|\Psi^s_{i,s}\rangle\langle \Psi^s_{i,s} | \op{S}^\dagger_{\lambda_m} \op{N}_z^\dagger}{2s+1} ,\label{eqn:rholambda}
\end{equation}
where $\op{N}_z = 2^s/\sqrt{\binom{2s}{s-\op{S}_z}(2s+1)}$. When applying the proposed Slater search algorithm to the pure state $\op{\rho}^{\lambda_m}_i$ of the TMOs, the maximal overlap is always obtained for $\op{S}_{\lambda_m}|\Psi^s_{i,s}\!'\rangle$, where $|{\Psi^s_{i,s}}\!'\rangle$ is the closest separable state to $|\Psi^s_{i,s}\rangle$. Hence the separable state
\begin{equation}
\op{\rho}_i' = \frac{1}{2s+1}\sum_{m = -s}^{s}\op{S}_{\lambda_m}|{\Psi^{s}_{i,s}}\!'\rangle\langle{\Psi^{s}_{i,s}}\!'|\op{S}^\dagger_{\lambda_m},\label{eq:rhoprime}
\end{equation}
gives a local minimum of $E_G[\op{\rho}^{T}_i]$. Moreover each term $\op{\rho}^{\lambda_m}_i$ contains the same amount of entanglement, originating from the material specific $|\Psi^s_{i,s}\rangle$ and the action of the renormalization operator $\op{N}_z$. Therefore we conjecture that the local minimum given by $\op{\rho}_i'$ is in fact a global minimum. The geometric entanglement measure for the PM state $\op{\rho}^T$ can then be written as~\cite{edsupplemental}
\begin{equation}
E_G[\op{\rho}^T] = 1 - \frac{1-E_G[\op{\rho}^T_s]}{2^{2s}(2s+1)} \left[\sum_{m=-s}^s \sqrt{{\binom{2s}{s-m}}}\right]^2,\label{eqn:eft}
\end{equation}
where $\op{\rho}^T_s = \sum_i p_i |\Psi^{s}_{i,s}\rangle\langle\Psi^{s}_{i,s}|$. Even when $E_G[\op{\rho}^T_s]$ is zero there is still a non-trivial term remaining in Eq. (\ref{eqn:eft}). This source of entanglement is very robust as it does not depend on the details of the electronic structure of the system but only on $\langle\op{S}^2\rangle$.

As a complement to the geometric entanglement measure we have used an entropic measure, based on the reduction of the many-body density matrix $\op{\rho}$ to the one-particle reduced density matrix $\oop{\rho}_{ij} = \Tr(\op{\rho}c^\dagger_j c_i)$. This reduction converts the entanglement in $\op{\rho}$ to entropy, which can be measured directly~\cite{amico08rmp80:517,lowdin55pr97:1474}, e.g. in form of linear entropy $S_{L}[\oop{\rho}] = \Tr[\oop{\rho}(\oop{1}-\oop{\rho})]$. However, quantifying entanglement in form of entropy requires care, as also the classical correlation in $\op{\rho}$ is converted to entropy. We propose the following entropic entanglement measure
\begin{equation}
E_{L}[\op{\rho}] = S_{L}[\oop{\rho}] - \min\big(\Tr[\oop{\rho}],\Tr[\oop{1}-\oop{\rho}]\big)S_{L}[\op{\rho}]\label{eqn:el},
\end{equation}
based on the fact that the entropy gained from the classical correlations is less than the number of electrons or holes times the entropy in $\op{\rho}$~\cite{edsupplemental}. Removing this upper bound of the classical contribution simplifies the evaluation of $E_L$, but at the same time makes it less sensitive to detect entanglement in mixed states.

Apart from the entanglement measures proposed above, other choices are possible. An intermediate scheme is to evaluate the geometric entanglement by using the relative von Neumann entropy~\cite{vedral97prl78:2275}. However, the current lack of an efficient minimization procedure for the relative entropy with respect to the separable reference state $\op{\rho}'$ reduces its applicability to setting upper bounds on the entanglement~\cite{byczuk12}.

\begin{table*}
\caption{Entanglement of the thermal ground state at $T = 273 K$. The values of $s$ is defined by $s(s+1) = \langle\op{S}^2\rangle$, and the degenerate eigenstate $|\Psi_{m_s}\rangle$ corresponds to $m_s = -s,..,s$ for each value of $\langle\op{L}_z\rangle$. E (eV) is the relative energy of $|\Psi_{m_s}\rangle$, and $E_G$ and $E_L$ are defined in the text.\label{tab:ent}}
\begin{tabular}{ c c c c c c c c  c c c c c c @{\hspace{0.5cm}} c c c c c c }
\hline
\hline
TMO & $E_G[\op{\rho}_s^T]$ & $E_G[\op{\rho}^T]$ & $E_{L}[\op{\rho}^T]$ & E (eV) & $s$ & $\langle\op{L}_z\rangle$ &&  \multicolumn{6}{c}{$E_G[|\Psi^s_{m_s}\rangle\langle\Psi^s_{m_s}|]$} &  \multicolumn{6}{c}{$E_{L}[|\Psi^s_{m_s}\rangle\langle\Psi^s_{m_s}|]$} \\
      &      &      &       &      &     &            & $m_s=$ & 0    & $\pm 1/2$ & $\pm 1$ & $\pm 3/2$ & $\pm 2$ & $\pm 5/2$ & 0    & $\pm 1/2$ & $\pm 1$ & $\pm 3/2$ & $\pm 2$ & $\pm 5/2$ \\
\hline
MnO   & 0.00 & 0.15 & -1.67 & 0.00 & 5/2 & $~0.00$    &&      & 0.65      &         & 0.59      &         & 0.00      &      & 2.40      &         & 1.60      &         & 0.00      \\
FeO   & 0.00 & 0.11 & -1.40 & 0.00 & 2   & $~0.00$    && 0.62 &           & 0.58    &           & 0.00    &           & 2.01 &           & 1.51    &           & 0.01    &           \\
      &      &      &       & 0.00 & 2   & $\pm 0.98$ && 0.62 &           & 0.58    &           & 0.00    &           & 2.01 &           & 1.51    &           & 0.01    &           \\
CoO   & 0.02 & 0.09 & -0.81 & 0.00 & 3/2 & $~0.00$    &&      & 0.63      &         & 0.16      &         &           &      & 1.64      &         & 0.55      &         &           \\
      &      &      &       & 0.00 & 3/2 & $\pm 1.45$ &&      & 0.62      &         & 0.12      &         &           &      & 1.57      &         & 0.43      &         &           \\
NiO   & 0.00 & 0.03 & -0.35 & 0.00 & 1   & $~0.00$    && 0.50 &           & 0.00    &           &         &           & 1.00 &           & 0.00    &           &         &           \\
      &      &      &       & 0.08 & 1   & $~0.00$    && 0.50 &           & 0.00    &           &         &           & 1.00 &           & 0.00    &           &         &           \\
      &      &      &       & 0.08 & 1   & $\pm 0.49$ && 0.63 &           & 0.25    &           &         &           & 1.38 &           & 0.75    &           &         &           \\ 
\hline
\hline
\end{tabular}
\end{table*}
The many-body eigenstate decomposition of $\op{\rho}^T$ from the ED solver is shown in Table \ref{tab:ent}. The auxiliary bath orbitals are assigned a spin, but zero orbital angular momentum. The ground state of NiO contains two sets of near degenerate states, with an energy difference of 80 meV. Although the energy difference is small, the low temperature strongly favours the eigenstates with the lowest energy. In all the studied materials, the ground state configurations maximize $\langle\op{S}^2\rangle$ due to the strong Coulomb interaction.

We start by looking at the entanglement in the $\langle\op{S}_z\rangle$ and $\langle\op{L}_z\rangle$ resolved components $|\Psi^s_{m_s}\rangle\langle\Psi^s_{m_s}|$ of $\op{\rho}^T$. As seen in Table \ref{tab:ent}, both $E_G$ and $E_L$ increase in magnitude as $|m_s|$ becomes smaller. This trend follows the number of possible ways to distribute the electrons according to their spin~\cite{edsupplemental}. The entanglement in the maximally spin-polarized eigenstate $|\Psi^s_{s}\rangle$ depend mainly on the complicated interplay between the Coulomb interaction and the crystal field energies. The maximally spin-polarized ground state of NiO is dominated by a Ni $d^8$ high spin configuration ($\uparrow^2_3\downarrow^0_3$ where the subscript and the superscript stand for the number of t$_{2g}$ and e$_g$ electrons respectively). The on-site Coulomb interaction must preserve both $\langle\op{S}_z\rangle$ and $\langle\op{L}_z\rangle$ which implies that it can not couple this state to any other state. The Coulomb interaction is therefore effectively reduced to a Hartree-Fock term, which makes the eigenstate $|\Psi^s_{s}\rangle$ separable. The eigenstates at 80 meV have mainly $\uparrow^2_3\downarrow^1_2$ character. Here the Coulomb interaction is allowed to transform these states to $\uparrow^2_3\downarrow^2_1$ through pair-hopping. Nevertheless, when the maximally spin-polarized states are combined into the density matrix $\op{\rho}^T_s$ the entanglement in the states cancels out, giving  $E_G[\op{\rho}^T_s] = 0.00$.

The ground state of CoO has mainly a Co $\uparrow^2_3\downarrow^0_2$ configuration that couples to $\uparrow^2_3\downarrow^1_1$ through the pair-hopping induced by the Coulomb interaction. The pair-hopping gives rise to a strong entanglement in $|\Psi^s_{s}\rangle$, and in contrast to NiO, the entanglement is still present in $\op{\rho}^T_s$, giving $E_G[\op{\rho}^T_s] = 0.02$.

The ground state of FeO has primarily a Fe $\uparrow^2_3\downarrow^0_1$ configuration. The Coulomb interaction is again reduced to a Hartree-Fock term, except when the bath introduces an extra electron in the TM-3d orbitals. As a result the strength of the pair-hopping becomes very small, and it gives rise only to a very weak entanglement with $E_G[|\Psi^s_{s}\rangle\langle\Psi^s_{s}|] = 0.003$ and $E_L[|\Psi^s_{s}\rangle\langle\Psi^s_{s}|] = 0.012$.

The MnO ground state is dominated by the Mn $\uparrow^2_3\downarrow^0_0$ configuration. Even when one extra electron is introduced from the bath, the Coulomb interaction cannot induce any pair hopping, which makes $\op{\rho}^T_s$  fully separable.

The Coulomb interaction gives an additional contribution to the entanglement when two electrons are transferred from the bath to the TM-3d orbitals due to the quadratic increase in repulsion energy. However, for TMOs this contribution is of the order of $10^{-4}$ for the geometric measure, i.e. too small to be seen in Table \ref{tab:ent}.

Let us now consider the entanglement in the thermal PM ground state $\op{\rho}^T$. For NiO, FeO and MnO the $E_G[\op{\rho}^T_s]$ term in Eq. (\ref{eqn:eft}) is zero, and only the $\langle \op{S}^2 \rangle$-dependent part contributes to the entanglement. A single Slater determinant method can therefore {\itshape{in principle}} obtain the maximally spin-polarized states $|\Psi^s_{i,s}\rangle$, and then reconstruct the ground states through Eq. (\ref{eqn:rholambda}). However, such an approach would not be adequate for CoO, as $E_G[\op{\rho}^T_s]$ is non-zero in this case.

The $\langle \op{S}^2 \rangle$-contribution causes $E_G[\op{\rho}^T]$ to increase monotonically from NiO to MnO. A similar trend can in fact also be seen in the relative entropy between the locally projected~\cite{edsupplemental} thermally mixed density operators in LDA and LDA+DMFT~\cite{byczuk12}. However, due to a strong reduction of the large number of available many-body states in the LDA solution to only a few in the LDA+DMFT solution, the size of the relative entropy reflects mainly the different degree of classical correlation in the two calculations. This trend in the relative entropy occurs in fact even when the LDA+DMFT ground state is replaced by the separable ground state of a corresponding Hartree-Fock-like LDA+U simulation.

The proposed $E_L$ measure was not able to resolve the entanglement in the PM phase, as shown by the negative values of $E_L[\op{\rho}^T]$ in Table \ref{tab:ent}. Further refinement of the subtraction of the classical contribution is needed to extend its applicability beyond pure states. One possibility would be to explicitly include the number of orbitals in the set of conditions~\cite{edsupplemental} used to define the upper bound of the classical contribution.

Finally we note that the PM results can be extrapolated to the type-II anti-ferromagnetic (AFM) phase at zero Kelvin, by reducing the thermal ground state to the zero energy eigenstate with the largest magnetic moment. For NiO, FeO and MnO these eigenstates are separable, which again makes them describable within theories working with a single Slater determinant, at least in principle. In CoO this state is rather entangled, with a fidelity entanglement of $0.12$, which once more puts an upper bound on the accuracy of the single Slater determinant methods for this material. Nevertheless, we would like to stress that even though the extrapolated AFM ground states of NiO, FeO and MnO are separable their excited states are in general entangled. This means that it is not possible to fully capture the excitation spectrum of the TMOs using the band picture given by single Slater determinant methods.

In conclusion, we have proposed a method able to separate classical correlations from entanglement in a material specific theoretical framework. The ground states of the strongly correlated late TMOs have been studied in detail, and two main sources of entanglement have been identified. In the PM phase the entanglement is dominated by a contribution proportional to the atomic $\langle{\op{S}^2}\rangle$. The second contribution, present also in the zero temperature AF phase, comes from the pair hopping induced by the on-site Coulomb interaction. This effect is suppressed by the presence of the crystal field splitting, but plays still an important role in CoO. We expect this behaviour to be common to a wide range of materials with similar symmetries, e.g. Co doped ZnO. It would be of great interest to study the role of the entanglement in more exotic systems as complex actinides or Fe based superconductors, and see how it relates to their unconventional properties.  

 We are grateful to the Swedish Research Council, Energimyndigheten (STEM), the KAW foundation and ERC (247062 - ASD), for financial support. Calculations have been performed at the Swedish national computer centers UPPMAX and NSC. We like to thank S. Garnerone and M. I. Katsnelson for useful discussions.

%

%

\newpage
\appendix
\section{Supplemental material}

In the following formulas we have followed the convention of using hats for many-body operators, e.g.  $\op{U}$, and tildes  for one-particle operators, e.g. ${\oop{\Delta}}$. A many-body operator $\op{B}$ can be obtained from a one-particle operator $\oop{B}$ as
\begin{equation}
\op{B} = \sum_{ij} \oop{B}_{ij} \op{c}^\dagger_i\op{c}_j.
\end{equation}

\section{Exact Diagonalization}
\subsection{Block diagonalization}\label{sec:block}
The Exact Diagonalization (ED) method solves the Anderson impurity problem by mapping it to a finite size problem, defined by the Hamiltonian 
\begin{equation}
\op{H}^{ED} = \op{H}^0 + \op{U},
\end{equation}
where $\op{H}^0$ contains the one-electron terms, including the hybridization with a few auxiliary bath orbitals, and $\op{U}$ describes the full Coulomb interaction between the electrons in the correlated orbitals~\cite{caffarel94prl72:1545}. 

An important technical issue in the implementation of ED is constructing the matrix representation of the Hamiltonian with respect to the many-body basis. This step is greatly simplified if the basis vectors are defined as single Slater determinants in some one-particle basis. The Slater determinants can be visualized in the occupation number formalism. For example 5 electrons in a 10 orbital manifold can form the following many-body states:
\begin{eqnarray*}
|\Psi_1^5\rangle & = & | 1 1 1 1 1 0 0 0 0 0 \rangle,\\
|\Psi_2^5\rangle & = & | 1 1 1 1 0 1 0 0 0 0 \rangle,\\
|\Psi_3^5\rangle & = & | 1 1 1 1 0 0 1 0 0 0 \rangle,\\
& \vdots & \\
|\Psi_{M-1}^5\rangle & = & | 0 0 0 0 1 0 1 1 1 1 \rangle,\\
|\Psi_M^5\rangle & = & | 0 0 0 0 0 1 1 1 1 1 \rangle.
\end{eqnarray*}
Here M is the number of possible many-body states, i.e. the binomial coefficient of 10 over 5. When the creation and annihilation operators are given in the same one-particle basis as $|\Psi^N_i\rangle$, then their action becomes a simple remapping of the indices $N \rightarrow N\pm1, i \rightarrow j$ and a multiplication of a phase factor ($\pm 1$). 

In a system with a large Hubbard U and bath energies $E$ far from the Fermi energy, it is usually sufficient to consider the ground-state configurations with $N$ electrons, and the excited configurations with $N\pm1$ electrons. The size of the many-body basis is then equal to all possible configurations of $N$, and $N\pm 1$ electrons distributed in $K$ one-electron spin-orbitals. However, even with a moderate number of spin-orbitals ($K = 30$) and close to complete filling ($N = 0.8 K = 24$), the size of the Hilbert space becomes too large to handle in practice. Nevertheless often the system under analysis possesses useful symmetries, which can be used to find some criteria to {\em{}a priori} identify a block structure in the Hamiltonian, and treat these blocks separately. 

A general way to obtain these criteria is to define the many-body basis vectors as the eigenvectors of a set of some commuting observables $\{\op{A}^k\}$, and then determine how the states mix under the action of the Hamiltonian. Since the basis vectors should be representable by single Slater determinants, only commuting one-electron observables
\begin{equation}
\op{A}^k = \sum_{ij} \oop{A}^k_{ij} \op{c}^\dagger_i\op{c}_j,
\end{equation}
need to be considered. Furthermore the additional condition that the observables should commute with $\op{U}$ allows to keep the calculation of the mixing rather simple. These two restrictions reduce the list of potential observables $\{\op{A}^k\}$ for the correlated orbitals to $\op{S}_z$ and $\op{L}_z$ (or alternatively $\op{S}_x$ and $\op{L}_x$ and $\op{S}_y$ and $\op{L}_y$). The auxiliary bath orbitals are not directly affected by $\op{U}$, so each bath spin-orbital $m$ can be assigned an observable $\op{n}^m = \op{c}^\dagger_m\op{c}_m$ that measures its occupation.

Each spin-orbital $j$ can be transformed into a common eigenstate of all these observables and assigned a vector of eigenvalues $\vec{a}^j$. A many-body basis vector $|\Psi^N_i\rangle$, defined as a single Slater determinant with respect to these orbitals, is trivially an eigenstate of the observables in $\{\op{A}^k\}$, with an eigenvalue vector
\begin{equation}
\vec{A}^i = \sum_{j=1}^K \langle\Psi^N_i|\op{c}^\dagger_j\op{c}_j|\Psi^N_i\rangle \vec{a}^j.
\end{equation}
Obtaining these eigenvalue vectors is not enough to determine the block structure of $\op{H}^{ED}$, as the one-electron Hamiltonian $\op{H}^0$ does not in general commute with the observables in $\{\op{A}^k\}$. In particular, the hybridization with the bath orbitals rarely commute with $\oop{L}_z$. The off-diagonal elements of $\oop{H}^0$ determine how the blocks will form. Each non-zero off-diagonal element $\oop{H}^0_{mn} \op{c}^\dagger_m\op{c}_n$ allows a many-body basis vector $|\Psi_i\rangle$ to couple to a basis vector $|\Psi_j\rangle$ if the following condition is fulfilled
\begin{equation}
\vec{A}^j - \vec{A}^i = \vec{a}^m - \vec{a}^n.\label{eqn:dmn}
\end{equation}
The problem of generating a block structure can now be transformed into finding the connected components of an undirected graph, where the vertices are defined by $\{\vec{A}^i\}$ and the edges by $\{\vec{a}^m - \vec{a}^n,0\}$ through Eq. (\ref{eqn:dmn}). This problem can be solved efficiently through the use of sparse logical square matrix multiplication in the following steps:
\begin{enumerate}
\item Map each vertex to a matrix index, and each edge to a true element in the logical matrix $T$.
\item Multiply $T$ with itself, until it remains constant. This procedure makes the connected components complete.
\item The true elements in a row or column in $T$ gives the indices of all the vertices of a connected component.
\end{enumerate}
This algorithm can be improved by contracting dense regions of the graph before $T$ is defined. Furthermore, the second step can be performed even more efficiently through the use of an update matrix $V$:
\begin{enumerate}
\item $V \leftarrow T$
\item While $V \neq 0$ 
\begin{enumerate}
\item $T \leftarrow T$ .or. $V$
\item $V \leftarrow$ (.not. $T$) .and. ($T \cdot V$)
\end{enumerate}
\end{enumerate}
The Hamiltonian is now finally ready to be block diagonalized by grouping all the many-body basis vectors with quantum number vectors $\{\vec{A}^i\}$ matching a given connected component.

\subsection{Correlated orbitals and the hybridization function}
The ED method is built around fitting a few hybridization parameters $\oop{V}$ and $\oop{E}$ to the hybridization function
\begin{equation}
\oop{\Delta}(\omega) = \omega\oop{1} - \oop{H}^{LDA-DC} - \left[\oop{G}^0(\omega)\right]^{-1},\label{eqn:hybr}
\end{equation}
where $\oop{G}^0(\omega)$ is the bath Green's function and $\oop{H}^{LDA-DC}$ is the projected LDA Hamiltonian with the double counting removed. In our implementation the hybridization function is given on the Matsubara axis $\omega = i\omega_n = i\pi T (2n + 1)$, and the fitting is performed by minimizing the cost function
\begin{equation}
F(\oop{V},\oop{E}) = \sum_n W_n\left\| \oop{V}^\dagger [i\omega_n\oop{1} - \oop{E}]^{-1} \oop{V} - \oop{\Delta}(\omega) \right\|_F^2,\label{eqn:fd}
\end{equation}
through the conjugate gradient method. Here $\{W_n\}$ is a set of weights and $\|\oop{A}\|_F^2 = \Tr(\oop{A}^\dagger\oop{A})$ is the Frobenius norm. For the ED method to give physically relevant results it is important that the original hybridization function can indeed be approximated by a small number of peaks. 

\begin{figure}
\includegraphics[width=8cm]{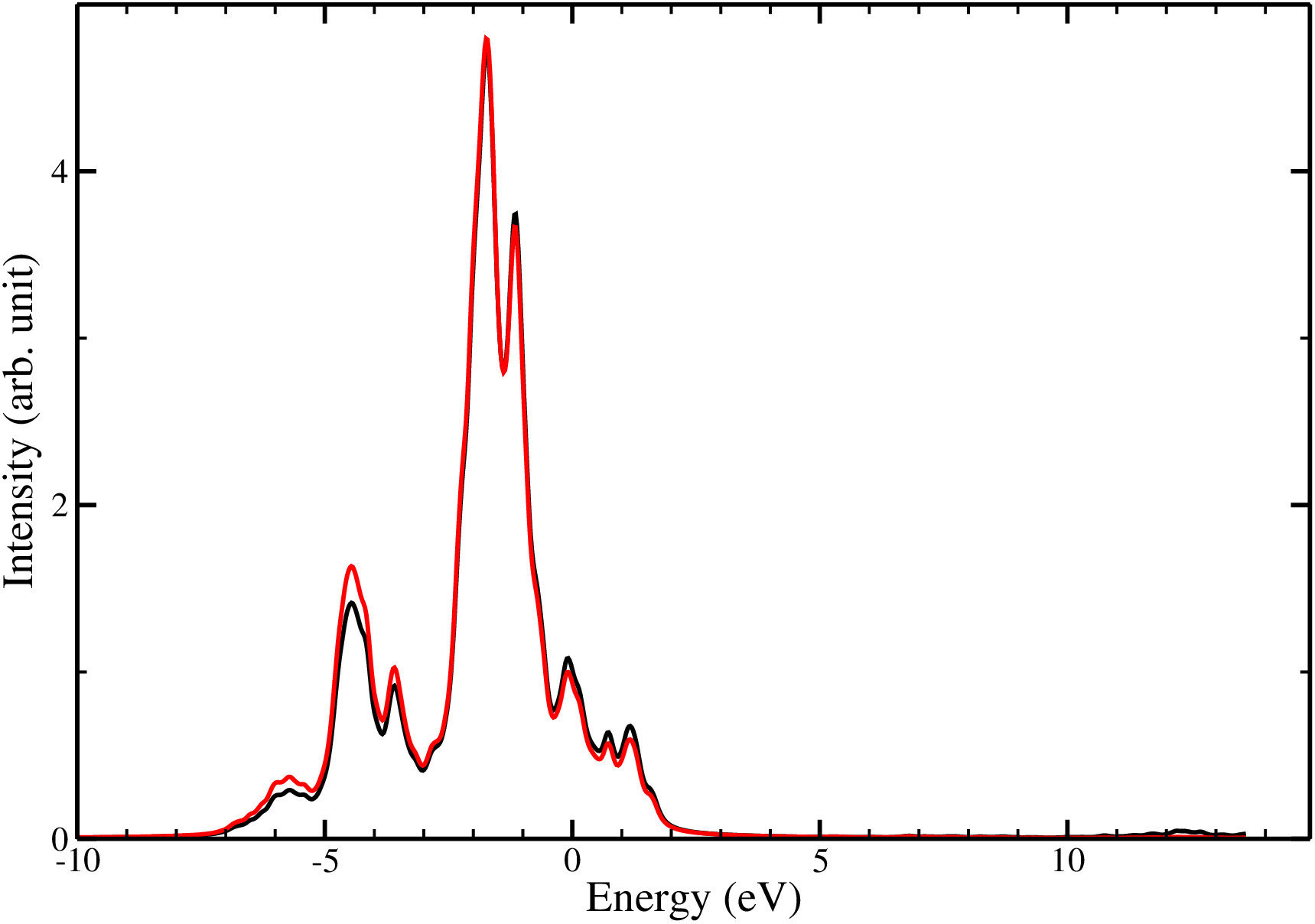}\\
\includegraphics[width=8cm]{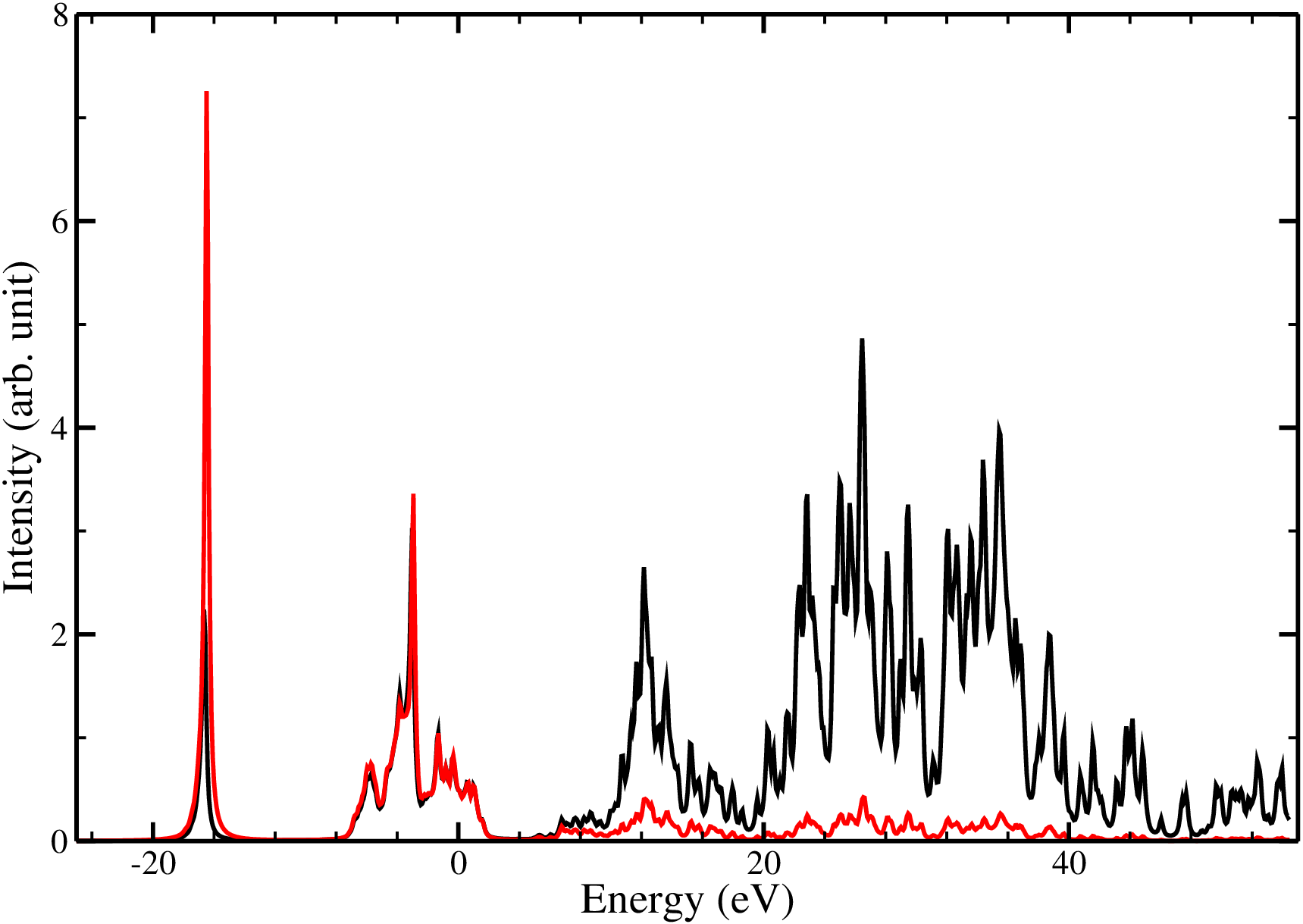}
\caption{(Colour online) Upper panel: The Ni-3d projected density of states of NiO in LDA. The black lines correspond to the MT orbitals and the red lines to the ORT orbitals defined in the text. Lower panel: Corresponding spectral density of the Ni-3d hybridization function in the same notation as above. \label{fig:nio}}
\end{figure}

The choice of correlated orbitals determines the mapping from the lattice problem to the effective impurity model, and therefore the value of $\oop{\Delta}(\omega)$. In order for the mapping to be realistic, the orbitals should be localized and nearly dispersionless. In the ED solver it is advantageous if the orbitals are eigenstates of the operator $\oop{L}_z$, as explained in previous section. Our current LDA+DMFT implementation~\cite{RSPt_book,grechnev07prb76:035107} contains two choices of orbitals that meet these conditions to a large extent, the heads of the LMTO's (MT) and L\"owdin orthogonalized  LMTO's (ORT). The MT orbitals are given by 
\begin{equation}
\psi_{lm}^{MT}(\vec{r}) = \phi_l(r) \theta(S - r) Y_{lm}(\hat{r}),
\end{equation}
where $\theta$ is a step function, $Y_{lm}$ is a spherical harmonic function, and $\phi_l(r)$ is the solution to the radial Schr\"odinger equation within the muffin-tin sphere of radius $S$. The MT orbitals are both localized and eigenstates of $\oop{L}_z$ by construction. The ORT orbitals at a site $\vec{R}$ are defined as
\begin{equation}
\psi_{l_i m_i}^{ORT}(\vec{r}) = \sum_{\vec{k}\in BZ}\sum_{j} e^{i\vec{k} \cdot \vec{R}} \psi_{\vec{k}j}^{LMTO}(\vec{r}) [O_k^{-1/2}]_{ji},
\end{equation}
where $\psi_{\vec{k}j}^{LMTO}(\vec{r})$ is an LMTO orbital, $O$ is the LMTO overlap matrix, the index $j$ runs over all the LMTO orbitals, and $\vec{k}$ runs over all the k-points in the Brillouin zone. The ORT orbitals are in general less localized and only approximately eigenstates to $\oop{L}_z$, due to the inclusion of the LMTO tail functions and the mixing through $O^{-1/2}$. However, when  $\psi_{l_i m_i}^{ORT}(\vec{r})$ correspond to localized states, both choices lead to similar orbitals. As an example, in the upper panel Fig. \ref{fig:nio} the Ni-3d projected density of states of NiO in LDA is reported. The densities for the MT and ORT bases are practically identical. Nevertheless, the small differences in the projections do have a large impact on $\oop{\Delta}$. In the lower panel of  Fig. \ref{fig:nio} we show the corresponding spectral density of the hybridization function
\begin{equation}
N^{hyb}(E) = -\frac{1}{\pi}\Im(\Tr[\oop{\Delta}(E+i\delta)]) ,
\end{equation}
in ORT and MT. A striking difference between them is that the MT orbitals give rise to a huge high-energy peak structure starting at 10 eV, while this is nearly absent when the ORT orbitals are used. Such feature is common to all the transition metals oxides of the present study - NiO, CoO, FeO, MnO - and to many other systems as well.
The strong high-energy hybridization given by the MT orbitals is very detrimental to the fitting process, since it attract the poles of the fitted function in Eq. (\ref{eqn:fd}). Therefore even though the ORT orbitals are less localized and require the use of a simplified basis~\cite{grechnev07prb76:035107,RSPt_book}, they are still to be preferred. The use of the ORT orbitals improves the situation, but it does not completely eliminate the high-energy hybridization. 

The weights $W_n$ can be chosen to make the cost function focus more on the physically relevant energy scale. In the calculations presented in the next section, $W_n$ was set to sample a logarithmic mesh from 7 to 200 eV. The initial tight spacing of the mesh points gives more weight to the residues on the eV scale, while the sparsely spaced high-energy points make the cost function smoother and increase the stability of the procedure. The Matsubara frequencies smaller than 7 eV were excluded since they biased the fitting procedure in favour of a detailed description close to the Fermi energy, at the expense of the overall description on the eV scale. 

The weights improve the fitting of the hybridization function on the Matsubara axis, but further refinement of the fitting procedure is necessary to obtain a good physical description for real energy values. To this end we used six bath states per spin-orbital in Eq. (\ref{eqn:fd}), but only the physically relevant elements with the largest value of the utility function 
\begin{equation}
M(\oop{V},\oop{E}) = -\frac{\oop{E}}{\oop{1}+3|\oop{E}|^3}\oop{V}\oop{V}^\dagger,
\end{equation}
were included in the final ED calculation. Here $M(\oop{V},\oop{E})$ is given in a basis where $\oop{E}$ is diagonal. The utility function $M(\oop{V},\oop{E})$ excludes contributions from the broad high-energy structure described above, and improves the stability with respect to an initially metallic density.

The post-selection procedure on the basis of the utility function worsens the fit on the Matsubara axis, but improves it closer to the real energy axis. It is important that a reasonable description of the original hybridization is obtained for both axes, and such situation is usually realized when the discarded states are well separated from the selected bath states.

\section{Computational details}\label{sec:comp}
The TMOs were all set up in the NaCl crystal structure. The lattice constants were taken from experimental data, and the Brillouin zone was sampled through a conventional Monkhorst-Pack mesh of 12 x 12 x 12 k-points. The calculations were carried out in the paramagnetic phase, using a finite temperature fully charge self-consistent LDA+DMFT implementation~\cite{grechnev07prb76:035107,dimarco09prb79:115111,thunstrom09prb79:165104,granas12cms55:295} based on the full-potential linear muffin-tin orbitals (LMTO) method~\cite{RSPt_book}. The double counting correction was defined as $\op{H}^{DC} = \mu^{DC}\op{N}$, where $\op{N}$ gives the number of electrons in the correlated orbitals. $\mu^{DC}$ was treated as a free parameter to ensure that the system has the correct number of electrons in the ground-state and that the chemical potential is placed in the band gap according to the experimental photoemission spectra. 

The Coulomb interaction was parametrized by the Slater parameters $F^0$, $F^2$, and $F^4$. The latter two were recalculated at every iteration from radial integration of the bare Coulomb interaction and then multiplied with the screening factors 0.82 and 0.88, respectively. The screening factors were set to reproduce the RPA screened Slater parameters for NiO in Ref. \onlinecite{kutepov10prb82:045105}. The parameter $F^0$, i.e. the Hubbard U, could not be treated in the same way as it is too strongly affected by the screening, but was set to a fixed value from the start. The $F^0$ values for MnO, FeO, CoO, and NiO were set to follow a linear increase in the local Coulomb interaction as the 3d-orbitals contract through the late transition metal series. The final self-consistent values of the Slater parameters are shown in Table \ref{tab:ed}. 

The ED calculations of NiO, CoO and FeO were performed with 10 TM 3d spin-orbitals and 20 auxiliary bath spin-orbitals. For MnO only 10 bath spin-orbitals were used, due to the greater computational effort. The final energy and hybridization strength parameters for the bath states are reported in Table \ref{tab:ed}. Note that the values are given in the crystal field basis, e$_g$ and t$_{2g}$, in which the hybridization function is diagonal. The auxiliary bath states between -17.9 and -19.5 eV correspond closely to the O 2s orbitals, while the rest have mainly O 2p character. 
\begin{table}
\caption{Self-consistent Slater parameters and auxiliary bath state fitting parameters. $F^0$ was held fixed at the tabulated value, while $F^2$, $F^4$, and the bath state parameters (Bath par.) were recalculated at each new iteration. The bath state parameters are given in the crystal field basis e$_g$ and t$_{2g}$ in which the hybridization function is diagonal. All values are given in eV.\label{tab:ed}}
\begin{tabular}{| l | c | c | c | c c | c c | c c | c c |}
\hline
 & \multicolumn{3}{c|}{Slater par.} &\multicolumn{4}{c|}{Bath par. (e$_g$)} & \multicolumn{4}{c|}{Bath par. (t$_{2g}$)} \\
\hline
TMO & F$^0$ & F$^2$ & F$^4$ & $E$ & $V$ & $E$ & $V$ & $E$ & $V$ & $E$ & $V$ \\ 
\hline
MnO & 6.0 & 9.0 & 6.1 & -5.1 & 2.1 & -- & -- & -6.2 & 1.4 & -- & -- \\
FeO & 6.5 & 9.2 & 6.2 & -4.4 & 1.9 & -17.9 & 2.1 & -8.9 & 0.7 & -4.6 & 1.2 \\
CoO & 7.0 & 10.0 & 6.7 & -4.6 & 1.8 & -17.9 & 2.0 & -7.5 & 0.9 & -4.0 & 1.0 \\
NiO & 7.5 & 10.1 & 6.7 & -5.7 & 1.9 & -19.5 & 2.0 & -8.8 & 0.9 & -5.1 & 0.8 \\
\hline
\end{tabular}
\end{table}

\section{Entanglement}
\subsection{Correlation and projection}
The term correlation has come to represent all electronic or quasi-particle interactions that go beyond a simple single Slater determinant description. It can be divided into two parts, classical and quantum correlation. The former gives rise to mixed states, and the latter to entangled states. The four different combinations of classical and quantum correlation result in the following classifications of an N-electron state in terms of wave-functions and density operators
\begin{equation*}
\begin{array}{l l}
\left.\begin{array}{c c l l}
|\Psi_i'\rangle & = & \displaystyle \prod_p^N \op{c}^\dagger_{i_p} |0\rangle\qquad\quad & {\mathrm{pure~separable}} \\
\end{array}~~~ \right\} &{\mathrm{Uncorr.}}\\
\left.\begin{array}{c c l l}
|\Psi_i\rangle & = & \displaystyle \sum_j a_{ij}\prod_p^N \op{c}^\dagger_{j_p} |0\rangle & {\mathrm{pure~entangled}} \\
\op{\rho}' & = & \displaystyle \sum_i p_i' |\Psi_i'\rangle\langle\Psi_i'| & {\mathrm{mixed~separable}} \\
\op{\rho} & = & \displaystyle \sum_i p_i |\Psi_i\rangle\langle\Psi_i| & {\mathrm{mixed~entangled}}
\end{array} \right\} & {\mathrm{Corr.}}
\end{array}
\end{equation*}
where $|0\rangle$ is the vacuum state. Here $p_i$ and $p_i'$ are sets of probabilities normalized to one, and $a_{ij}$ are expansion coefficients. The vectorial indices $i$ and $j$ are composed of the ordered components $i_p$ and $j_p$ referring to a generic one-particle basis. For example the state $\op{c}_1^\dagger \op{c}_3^\dagger \op{c}_7^\dagger |0\rangle$ corresponds to a many body index $i=(1,3,7)$. 

Classical correlations can derive from physical processes, e.g. thermal decoherence, but also be induced by a local projection upon some set of orbitals, e.g. the d- or f-orbitals of single atom in a solid. To see this, let us consider a set orbitals $A$ labelled by the indices $\{{j_p}\}_A$, and define the operator subspaces
\begin{eqnarray*}
\mathcal{A}' & = & \left\{ \op{A}^\dagger = \prod_p^{N_j} \op{c}^\dagger_{j_p}; j_p \in \{{j_p}\}_A \right\},\\
\mathcal{B}' & = & \left\{ \op{B}^\dagger = \prod_p^{N_j} \op{c}^\dagger_{j_p}; j_p \notin \{{j_p}\}_A \right\}.
\end{eqnarray*}
The full Hilbert space $\mathcal{H}$ can be partitioned as
\begin{equation*}
\mathcal{H}=\mathcal{A} \otimes \mathcal{B} \equiv  \left\{\sum_{ij} \alpha_{ij}\op{A}_i^\dagger\op{B}_j^\dagger|0\rangle;\op{A}_i^\dagger \in \mathcal{A}',\op{B}_j^\dagger \in \mathcal{B}'\right\} , 
\end{equation*}
where
\begin{eqnarray*}
\mathcal{A} &=& \left\{\op{A}^\dagger|0\rangle; \op{A}^\dagger \in \mathcal{A}' \right\},\\
\mathcal{B} &=& \left\{\op{B}^\dagger|0\rangle; \op{B}^\dagger \in \mathcal{B}' \right\} .
\end{eqnarray*}
The local projection of a pure state $|\Psi\rangle \in \mathcal{A} \otimes \mathcal{B}$ upon $A$ is obtained by taking the partial trace over $\mathcal{B}$:
\begin{equation}
\op{\rho}^A = \sum_{ij} \op{\rho}^A_{ij} \op{A}_i^\dagger|0\rangle\langle0| \op{A}_j,
\end{equation}
where $\op{\rho}^A_{ij} = \sum_k \alpha_{ik}\alpha^*_{jk}$. The locally projected state $\op{\rho}^A$ is in general a mixed state, unless $\op{\rho}^A_{ij}$ is idempotent. 

It becomes harder to distinguish two states after a local projection since some information is lost in the partial trace over $\mathcal{B}$. For example, the highly mixed state $\op{\rho}$, the pure and separable state $\op{\rho}'$, and the pure entangled state $\op{\rho}''$ given by
\begin{eqnarray*}
\op{\rho} & = & \frac{1}{4}\left(\op{c}^\dagger_1\op{c}^\dagger_2|0\rangle\langle0|\op{c}_2\op{c}_1 + \op{c}^\dagger_1\op{c}^\dagger_3|0\rangle\langle0|\op{c}_3\op{c}_1 + \right.\\
& & \quad \: \, \left. \op{c}^\dagger_2\op{c}^\dagger_4|0\rangle\langle0|\op{c}_4\op{c}_2 + \op{c}^\dagger_3\op{c}^\dagger_4|0\rangle\langle0|\op{c}_4\op{c}_3\right),\\
\op{\rho}' & = & \frac{\op{c}^\dagger_1+\op{c}^\dagger_4}{\sqrt{2}}\frac{\op{c}^\dagger_2+\op{c}^\dagger_3}{\sqrt{2}}|0\rangle\langle0|\frac{\op{c}_2+\op{c}_3}{\sqrt{2}}\frac{\op{c}_1+\op{c}_4}{\sqrt{2}},\\
\op{\rho}'' & = & \left({\op{c}^\dagger_1\frac{\op{c}^\dagger_2+\op{c}^\dagger_3}{\sqrt{2}}+\frac{\op{c}^\dagger_2-\op{c}^\dagger_3}{\sqrt{2}}\op{c}^\dagger_4}\right) |0\rangle \\
& & \langle0|\left({\frac{\op{c}_2+\op{c}_3}{\sqrt{2}}\op{c}_1+\op{c}_4\frac{\op{c}_2-\op{c}_3}{\sqrt{2}}}\right),
\end{eqnarray*}
become indistinguable after a local projection onto the orbitals 1 and 2 ($\{1,2\}_A$). It is therefore in general not possible to draw any conclusions about the classical or quantum correlations in the original unprojected state from its local projection.

Quantum correlations can stem from physical electron-electron interaction, e.g. Coulomb interaction, or be induced by a non-local projection. A non-local projection is qualitatively different from a local projection in that it projects onto a set of many-body states, and not a set of orbitals. The entanglement of the projected state may be larger than that of the original state. A trivial example is when the non-local projection is defined with respect to a single entangled state $|\Psi\rangle$, as any separable state with a non-zero overlap with $|\Psi\rangle$ becomes entangled after the projection. A less trivial example is given by the projection operator
\begin{equation}
\op{P} = \op{c}^\dagger_{1\uparrow}\op{c}^\dagger_{2\downarrow}|0\rangle\langle0|\op{c}_{2\downarrow}\op{c}_{1\uparrow} + \op{c}^\dagger_{1\downarrow}\op{c}^\dagger_{2\uparrow}|0\rangle\langle0|\op{c}_{2\uparrow}\op{c}_{1\downarrow}.
\end{equation}
The projection with $\op{P}$ makes the pure separable spin-coherent state $(1/2) (\op{c}^\dagger_{1\uparrow}+\op{c}^\dagger_{1\downarrow})(\op{c}^\dagger_{2\uparrow}+\op{c}^\dagger_{2\downarrow})|0\rangle$ entangled. In contrast to the first example it should be noted that $\op{P}$ projects onto two separable states. Moreover, the projection itself corresponds to a simple Stern-Gerlach experiment of a spin-1 particle with a single slit to remove the spin $\pm 1$ components.

\subsection{Pure separable state search algorithm}
The problem of finding the maximal overlap $O_{max}$ between an N-electron pure state $|\Psi\rangle$ and any pure separable state can be written as
\begin{equation}
O_{max}\left[|\Psi\rangle\right] = \max_{\oop{V}} \left|\langle 0|\sum_{i_1\cdots i_N}^K \op{c}_{i_N} \oop{V}_{i_N N} \cdots\op{c}_{i_1} \oop{V}_{i_1 1}  |\Psi\rangle\right|^2,\label{eqn:omax}
\end{equation}
where $K$ is the number of orbitals and $\oop{V}$ is a unitary matrix. A direct numerical optimization of the overlap, with respect to the parameters in $\oop{V}$, becomes quickly very cumbersome as the number of orbitals and electrons increase. A practical alternative to a brute force numerical optimization is the search function $F$, schematically presented in pseudocode form in Fig. \ref{fig:search}. The algorithm mimics an experiment where a sequence of one-electron removal measurements are performed. It is easy to visualize the search for the optimal $\oop{V}$ as an extremization of the intermediate intensities at each step in the measurement process. As the intensity monotonically decreases at each step, it is advantageous to cast the algorithm as a recursive depth first search function $F\left[|\Psi\rangle,O_{max}'\right]$, where $O_{max}'$ is the currently known maximal overlap. 
\begin{figure}
\hrule
~
\begin{flushleft}
{\bf function} $F$ is:\\
{\bf input:} pure state $|\Psi\rangle$, current maximal overlap $O_{max}'$
\begin{enumerate}
\item {\it \#Measurement optimization}:\\ Diagonalize the one-particle density matrix $\oop{\rho} = \oop{V} \oop{D} \oop{V}^\dagger$, where $\oop{\rho}_{ij} = \langle\Psi| \op{c}^\dagger_j \op{c}_i | \Psi\rangle$, such that the eigenvalues $\oop{D}_{ii}$ are sorted from largest to smallest.
\item {\it \#Detect the vacuum state:}\\ if ($\oop{D}_{11} = 0$) {\bf return} $\langle\Psi|\Psi\rangle$
\item {\it \#Perform the measurement and add another one-particle detector:}
\begin{tabbing}
fo\=r $i = 1,2,\cdots,$\mbox{ number of orbitals}\\
\> if \= ($\oop{D}_{ii} > O_{max}'$) then\\
\>\> $O_{max}' \leftarrow \max\left(O_{max}',F\left[\sum_j \op{c}_j \oop{V}_{ji} |\Psi\rangle,O_{max}'\right]\right)$\\
\> end if\\
done
\end{tabbing}
\item {\bf return} $O_{max}'$
\end{enumerate}
{\bf end $F$}
\end{flushleft}
\hrule
\caption{A pseudocode representation of the recursive depth first search function $F\left[|\Psi\rangle,O_{max}'\right]$. Note that the state $|\Psi\rangle$ is not renormalized after each measurement, and that the initial value of $O_{max}'$ can be set to zero. Comments are given in italic.\label{fig:search}}
\end{figure}

Although $F\left[|\Psi\rangle,0\right]$ may not return the maximal overlap for a general state, it is guaranteed to be exact if $|\Psi\rangle$ is separable or represents a 2-electron system. In the former case $|\Psi\rangle$ can be represented as a single Slater determinant. Its natural orbitals are therefore either completely occupied or empty, and the algorithm will return an overlap equal to one. The latter case can be proved by noting that the first electron removal leads to a state with one single electron, which is necessarily separable. Hence, no intensity will be lost in the second measurement, which implies that the maximal overlap is equal to the maximal probability of successfully removing the first electron. 

A reason the recursive search algorithm is not exact for systems with three or more electrons is that the optimal state is not always a superposition of pair-wise entangled states. This can be illustrated by applying a spin ladder operator to a maximally spin-polarized separable 3-electron state $| \Psi^{s=3/2}_{m=3/2} \rangle = \op{c}^{\dagger}_{\uparrow 1} \op{c}^{\dagger}_{\uparrow 2} \op{c}^{\dagger}_{\uparrow 3} |0\rangle$, giving
\begin{equation}
\frac{\op{S}^-}{\sqrt{3}} |\Psi\rangle = \frac{1}{\sqrt{3}}\left(\op{c}^{\dagger}_{\downarrow 1} \op{c}^{\dagger}_{\uparrow 2} \op{c}^{\dagger}_{\uparrow 3} - \op{c}^{\dagger}_{\uparrow 1} \op{c}^{\dagger}_{\downarrow 2} \op{c}^{\dagger}_{\uparrow 3} + \op{c}^{\dagger}_{\uparrow 1} \op{c}^{\dagger}_{\uparrow 2} \op{c}^{\dagger}_{\downarrow 3}\right)\!|0\rangle\label{eqn:sp3}
\end{equation}
The three Slater determinants on the right hand side in Eq. (\ref{eqn:sp3}) are all pair-wise entangled, which implies that the corresponding reduced one-particle density matrix $\oop{\rho}$ is diagonal. If the Slater search algorithm is applied to this state it will therefore keep the current basis and yield a maximal squared overlap of 1/3. However, the unitary spin transformation
\begin{equation}
\left(\begin{array}{c}
\op{c}'^{\dagger}_{\uparrow i} \\
\op{c}'^{\dagger}_{\downarrow i}
\end{array}\right) = \frac{1}{\sqrt{3}}
\left(\begin{array}{c c}
\sqrt{2} & 1 \\
-1 & \sqrt{2}
\end{array}\right) 
\left(\begin{array}{c}
\op{c}^{\dagger}_{\uparrow i} \\
\op{c}^{\dagger}_{\downarrow i}
\end{array}\right)\label{eqn:spoptimal} 
\end{equation}
gives
\begin{eqnarray}
\frac{\op{S}^-}{\sqrt{3}} |\Psi\rangle & = & \frac{2}{3} \op{c}'^{\dagger}_{\uparrow 1} \op{c}'^{\dagger}_{\uparrow 2} \op{c}'^{\dagger}_{\uparrow 3} |0\rangle + \frac{1}{3} \op{c}'^{\dagger}_{\uparrow 1} \op{c}'^{\dagger}_{\downarrow 2} \op{c}'^{\dagger}_{\downarrow 3} |0\rangle\nonumber\\
&& + \frac{1}{3}\op{c}'^{\dagger}_{\downarrow 1} \op{c}'^{\dagger}_{\uparrow 2} \op{c}'^{\dagger}_{\downarrow 3} |0\rangle - \frac{1}{3}\op{c}'^{\dagger}_{\downarrow 1} \op{c}'^{\dagger}_{\downarrow 2} \op{c}'^{\dagger}_{\uparrow 3} |0\rangle\nonumber\\
&& - \sqrt{\frac{2}{9}} \op{c}'^{\dagger}_{\downarrow 1} \op{c}'^{\dagger}_{\downarrow 2} \op{c}'^{\dagger}_{\downarrow 3} |0\rangle.\label{eqn:sptrans}
\end{eqnarray}
The first Slater determinant on the right hand side of Eq. (\ref{eqn:sptrans}) has a squared overlap of 4/9, which is larger than the overlap of 1/3 found for the untransformed basis. The spin transformation in Eq. (\ref{eqn:spoptimal}) is related to the number of ways the electrons can arrange their spins. For an arbitrary spin-resolved state $(\op{S}^-)^n |\Psi^s_s\rangle$ it takes the form
\begin{equation}
\left(\begin{array}{c}
\op{c}'^{\dagger}_{\uparrow i} \\
\op{c}'^{\dagger}_{\downarrow i}
\end{array}\right) = \frac{1}{\sqrt{a}}
\left(\begin{array}{c c}
\sqrt{a-b} & \sqrt{b} \\
-\sqrt{b} & \sqrt{a-b}
\end{array}\right) 
\left(\begin{array}{c}
\op{c}^{\dagger}_{\uparrow i} \\
\op{c}^{\dagger}_{\downarrow i}
\end{array}\right),\label{eqn:spgoptimal} 
\end{equation}
where $a = \binom{2s}{n}$ and $b = \binom{2s-1}{n}$. 

To check the accuracy of our results for the transition metal oxides we applied a numerical conjugate gradient (CG) scheme set to maximize the overlap in Eq. (\ref{eqn:omax}) directly via the parameters in the unitary transformation $\oop{V}$. The obtained maximal overlaps were all in exellent agreement with the results obtained from the search algorithm once the extra spin transformations in Eq. (\ref{eqn:spgoptimal}) had been applied to the spin-resolved states.

The search space of the algorithm can of course be extended further, e.g. by shifting the order of the electron removal measurements or using the final state as an input to a CG scheme, but such development is beyond the scope of this article.

\subsection{Entropic entanglement measure}
For a general many-body density operator $\op{\rho}$, the linear entropy
\begin{equation*}
S_{L}[\op{\rho}] = \Tr[\op{\rho}(\op{1}-\op{\rho})]
\end{equation*}
measures how far such a state is from being pure. It has been known for a long time~\cite{lowdin55pr97:1474} that $\op{\rho}$ can be represented by a single Slater determinant, i.e. a pure and separable state, if and only if the one-particle reduced density matrix $\oop{\rho}_{ij} = \Tr(\op{\rho} \op{c}^\dagger_j \op{c}_i)$ is idempotent. The linear entropy of the one-particle reduced density matrix
\begin{equation*}
S_{L}[\oop{\rho}] = \Tr[\oop{\rho}(\oop{1}-\oop{\rho})]
\end{equation*}
measures the deviation from idempotency, which makes it sensitive to both classical correlations and entanglement in $\op{\rho}$. A lower bound on the contribution from the entanglement can be obtained by removing the maximal contribution from a mixed but separable ensemble \mbox{$\op{\rho}' = \sum_{i=1}^Q p_i' |\Psi_i'\rangle\langle\Psi_i'|$}, under the constrain that \mbox{$S_L[\op{\rho}'] = S_L[\op{\rho}]$} and \mbox{$N = \Tr[\oop{\rho}] = \Tr[\oop{\rho}']$}. Let us consider a fictitious extended system, with $N$ electrons and $QN$ orbitals, for which we define
\begin{equation*}
\oop{\rho}'_{e} = \sum_i^Q p_i'\op{c}_{Ni}^\dagger \cdots \op{c}_{1+N(i-1)}^\dagger |0\rangle \langle0| \op{c}_{1+N(i-1)}^{\phantom\dagger} \cdots \op{c}_{Ni}^{\phantom\dagger}  .
\end{equation*}
The concavity of $S_L$ implies that 
\begin{equation*}
 S_L[\oop{\rho}'] \leq S_L[\oop{\rho}'_{e}].
\end{equation*}
Evaluating $S_L[\oop{\rho}'_e]$ gives
\begin{equation*}
S_L[\oop{\rho}'] \leq NS_L[\op{\rho}] = \Tr[\oop{\rho}]S_L[\op{\rho}]
\end{equation*}
An analagous construction can be made for \mbox{$M = \Tr[\oop{1}] - N$} holes and $QM$ orbitals, giving that
\begin{equation*}
S_L[\oop{\rho}'] \leq \Tr[\oop{1}-\oop{\rho}]S_L[\op{\rho}].
\end{equation*}
Hence, the state $\op{\rho}$ is necessarily entangled if
\begin{eqnarray}
E_{L}[\op{\rho}] & = & S_L[\oop{\rho}] - \min(\Tr[\oop{\rho}],\Tr[\oop{1}-\oop{\rho}])S_L[\op{\rho}]\nonumber\\
& = & S_L[\oop{\rho}] - \min(N,M)S_L[\op{\rho}]   > 0.
\end{eqnarray}
However, the reversed implication is not true in general. Any mixed state of the form 
\begin{equation*}
\op{\rho} = \frac{1}{Q} \sum_{i=1}^{Q} |\Psi_i\rangle\langle\Psi_i|,
\end{equation*}
where $\langle\Psi_i|\Psi_j\rangle = \delta_{ij}$ and $Q$ larger than the number of orbitals, gives $E_{L}[\op{\rho}] < 0$.

It should be noted that $E_{L}$ and the fidelity based entanglement measure
\begin{equation*}
E_F[\op{\rho}] \equiv 1 - \max_{\op{\rho}'} \Tr\left[\sqrt{\sqrt{\op{\rho}}\op{\rho}'\sqrt{\op{\rho}}}\right]^2,
\end{equation*}
where $\op{\rho}'$ is separable, are not equivalent even for pure states, in the sense there exist states $|\Psi_a\rangle$ and $|\Psi_b\rangle$ such that
\begin{eqnarray*}
E_F[|\Psi_a\rangle\langle|\Psi_a|] & < & E_F[|\Psi_b\rangle\langle|\Psi_b|],\\
E_{L}[|\Psi_a\rangle\langle|\Psi_a|] & > & E_{L}[|\Psi_b\rangle\langle|\Psi_b|].
\end{eqnarray*}
For example let us set
\begin{eqnarray*}
|\Psi_a\rangle & = & \frac{1}{\sqrt{2}}\left(\op{c}_1^\dagger\op{c}_2^\dagger + \op{c}_3^\dagger\op{c}_4^\dagger\right)  | 0 \rangle,\\
|\Psi_b\rangle & = & \left(\sqrt{x}\op{c}_1^\dagger\op{c}_2^\dagger + \sqrt{\frac{1-x}{2}}\op{c}_3^\dagger\op{c}_4^\dagger + \sqrt{\frac{1-x}{2}}\op{c}_5^\dagger\op{c}_6^\dagger \right) |0\rangle,
\end{eqnarray*}
then for $1/2 < x < 2/3$
\begin{eqnarray*}
E_F[|\Psi_a\rangle\langle|\Psi_a|] & = & \frac{1}{2} > 1-x = E_F[|\Psi_b\rangle\langle|\Psi_b|],\\
E_{L}[|\Psi_a\rangle\langle|\Psi_a|] & = & 1 < (1-x)(1+3x) = E_{L}[|\Psi_b\rangle\langle|\Psi_b|].
\end{eqnarray*}
 The difference comes from that $E_F$ only focuses on the largest overlapping pure separable state $|\Psi'\rangle$, while $E_L$ also takes into account the entanglement between the states orthogonal to $|\Psi'\rangle$.

\subsection{Entanglement and spin-coherence}
Any maximally spin-polarized N-electron state with spin quantum number $m_s = s$ can be decomposed as
\begin{equation}
|\Psi^s_{i,s}\rangle = \sum_j {a}_{ij} \prod_{q=1}^{2s} \op{c}^\dagger_{\uparrow j_q} \prod_{p=2s+1}^{N/2+s}\op{c}^\dagger_{\uparrow j_p}\op{c}^\dagger_{\downarrow j_p} |0\rangle,\label{eqn:p}
\end{equation}
where $|0\rangle$ is the vacuum state and ${a}_{ij}$ are the coefficients defined above. Applying the spin-coherent operator
\begin{equation}
\op{S}_\lambda = e^{\lambda\op{S}^-} (1+|\lambda|^2)^{-\op{S}_z}
\end{equation}
to $|\Psi^s_{i,s}\rangle$ gives
\begin{equation}
 \op{S}_\lambda |\Psi^s_{i,s}\rangle = \sum_j \op{a}_{ij} \prod_{q=1}^{2s} \frac{\op{c}^\dagger_{\uparrow j_q} + \lambda \op{c}^\dagger_{\downarrow j_q}}{\sqrt{1+|\lambda|^2}} \prod_{p=2s+1}^{N/2+s}\op{c}^\dagger_{\uparrow j_p}\op{c}^\dagger_{\downarrow j_p} |0\rangle, \label{eqn:sp}
\end{equation}
where we have used that 
\begin{equation}
e^{\lambda\op{S}^-} = \sum_{n=0}^{\infty} \frac{\left(\sum_i \lambda \op{c}^\dagger_{\downarrow i} \op{c}_{\uparrow i} \right)^n}{n!} . \label{eqn:ssum}
\end{equation}
The unitary transformation
\begin{equation}
\left(\begin{array}{c}
\op{c}'^{\dagger}_{\uparrow i} \\
\op{c}'^{\dagger}_{\downarrow i}
\end{array}\right) = \frac{1}{\sqrt{1+|\lambda|^2}}
\left(\begin{array}{c c}
1 & \lambda \\
-\lambda^* & 1
\end{array}\right) 
\left(\begin{array}{c}
\op{c}^{\dagger}_{\uparrow i} \\
\op{c}^{\dagger}_{\downarrow i}
\end{array}\right) 
\end{equation}
brings Eq. (\ref{eqn:sp}) back to exactly the same form as Eq. (\ref{eqn:p}), which proves that $\op{S}_\lambda$ conserves the entanglement of $|\Psi^s_{i,s}\rangle$. 

The 2s+1 roots of unity $\lambda_m = \exp[2i\pi m/(2s+1)]$ satisfy the important relation
\begin{equation}
 \sum_{m = -s}^s \lambda_m^{p}\lambda_m^{*q} = \sum_{m = -s}^s \exp\!\left(2i\pi \frac{m(p-q)}{2s+1}\right) = (2s+1)\delta_{pq},
\end{equation}
for all $-s \le p,q \le s$. Applying $\op{S}_{\lambda_m}$ to $|\Psi^s_{i,s}\rangle\langle\Psi^s_{i,s}|$ yields
\begin{eqnarray}
\op{\rho}_i &= & \sum_{m=-s}^{s} \frac{\op{S}_{\lambda_m} |\Psi^s_{i,s}\rangle\langle\Psi^s_{i,s}| \op{S}_{\lambda_m}^\dagger}{2s+1} \nonumber\\
&= & \sum_{m=-s}^{s} \!\left(\sum_{n=0}^{2s} \frac{\lambda_m^n (\op{S}^-)^n}{n!\, 2^s} \! \right) \! \frac{|\Psi^s_{i,s}\rangle\langle\Psi^s_{i,s}|}{2s+1} \! \left( \sum_{n=0}^{2s} \frac{\lambda_m^{*n} (\op{S}^+)^n}{ n!\, 2^s} \! \right)\nonumber\\
& = & \sum_{m,m_s,m_s'=-s}^{s}\!\!\!\! \op{N}_z^{-1} \lambda_m^{s-m_s} \frac{|\Psi^s_{i,m_s}\rangle\langle\Psi^s_{i,m_s'}|}{(2s+1)^2}  \lambda_m^{*s-m_s'} \op{N}_z^{-1} \nonumber\\
& = & \sum_{m_s = -s}^s  \frac{\op{N}_z^{-1} |\Psi^s_{i,m_s}\rangle\langle\Psi^s_{i,m_s}| \op{N}_z^{-1}}{2s+1},\label{eqn:rhoin}
\end{eqnarray}
where $\op{N}_z = 2^s/\sqrt{\binom{2s}{s-\op{S}_z}(2s+1)}$ and 
\begin{equation}
|\Psi^s_{i,m_s}\rangle = \frac{(\op{S}^-)^{s-m_s}}{(s-m_s)!\sqrt{\binom{2s}{s-m_s}}}|\Psi^s_{i,s}\rangle.\label{eqn:pms}
\end{equation}
Through Eq. (\ref{eqn:rhoin}) the thermal paramagnetic density operator 
\begin{equation}
\op{\rho}^T =\!\! \sum_{m_s=-s}^s \!\!\frac{\op{\rho}^T_{m_s}}{2s+1} \equiv \sum_{m_s=-s}^s \frac{ \sum_i p_i|\Psi^s_{i,m_s}\rangle\langle\Psi^s_{i,m_s}|}{2s+1}
\end{equation}
can  be written as
\begin{equation}
\op{\rho}^T = \!\sum_{m_s=-s}^s \frac{\op{N}_z\op{S}_{\lambda_m} \op{\rho}^T_{s} \op{S}_{\lambda_m}^\dagger\op{N}_z}{2s+1}.
\end{equation}

We can now introduce a set of maximally spin-polarized separable states $|\Psi^s_{i,s}\!'\rangle$ and use Eq. (\ref{eqn:pms}) to define
\begin{equation*}
\op{\rho}_{m_s}'  \equiv  \sum_i p_i |\Psi^s_{i,m_s}\!'\rangle\langle\Psi^s_{i,m_s}\!'|
\end{equation*}
and
\begin{equation*}
 \op{\rho}' \equiv \sum_{m = -s}^s \frac{\op{S}_{\lambda_{m}}|\Psi^s_{i,s}\!'\rangle\langle\Psi^s_{i,s}\!'| \op{S}_{\lambda_{m}}^\dagger}{2s+1} =  \! \sum_{m_s = -s}^s  \frac{\op{N}_z^{-1} \op{\rho}_{m_s}' \op{N}_z^{-1}}{2s+1}.
\end{equation*}
The entanglement preserving properties of $\op{S}_\lambda$ entails that $\op{\rho}'$ is separable as well. From Eq. (\ref{eqn:pms}) we also have that 
\begin{equation}
\langle\Psi^s_{j,m_s'}\!'|\Psi^s_{i,m_s}\rangle = \delta_{m_sm_s'} \langle\Psi^s_{j,s}\!'|\Psi^s_{i,s}\rangle,\label{eqn:ort}
\end{equation}
which implies that the fidelity between $\op{\rho}_{m_s}'$ and $\op{\rho}^T_{m_s}$ becomes spin-independent
\begin{equation}
F[\op{\rho}_{m_s}',\op{\rho}^T_{m_s}] \equiv \Tr\!\left[\sqrt{\sqrt{\op{\rho}^T_{m_s}}\op{\rho}'_{m_s}\sqrt{\op{\rho}^T_{m_s}}}~\right]^2 \!\!= F[\op{\rho}_{s}',\op{\rho}^T_{s}].\label{eqn:fnos}
\end{equation}
Substituting Eq.  (\ref{eqn:ort}) and  (\ref{eqn:fnos}) into $F[\op{\rho}',\op{\rho}^T]$ gives
\begin{eqnarray}
F[\op{\rho}',\op{\rho}^T] & = & \Tr\!\left[\sqrt{\sqrt{\op{\rho}^T}\op{\rho}'\sqrt{\op{\rho}^T}}~\right]^2\nonumber\\
& = & \Tr\!\left[\sqrt{\sum_{m =-s}^s \frac{\sqrt{\op{\rho}^T_m} \op{N}_z^{-1} \op{\rho}_m' \op{N}_z^{-1} \sqrt{\op{\rho}^T_m}}{(2s+1)^2}}~\right]^2\nonumber\\
& = & {\left(\sum_{m_s=-s}^s \sqrt{\frac{F[\op{\rho}^T_{m_s},\op{\rho}_{m_s}']\binom{2s}{s-m_s}}{2^{2s}(2s+1)}}\right)\!\!}^2 \nonumber\\
& = & \frac{F[\op{\rho}_s',\op{\rho}_s^T]}{2^{2s}(2s+1)}{\left(\sum_{m_s=-s}^s \sqrt{\binom{2s}{s-m_s}}\right)\!\!}^2.\label{eqn:frrt}
\end{eqnarray}
Finally, under the assumption that a separable density operator of the same form as $\op{\rho}'$ gives the maximal fidelity in Eq. (\ref{eqn:frrt}), the fidelity based entanglement measure can be written as
\begin{equation}
E_F[\op{\rho}^T] = 1 - \frac{1-E_F[\op{\rho}^T_s]}{2^{2s}(2s+1)}{\left(\sum_{m_s=-s}^s \sqrt{\binom{2s}{s-m_s}}\right)\!\!}^2.
\end{equation}

\end{document}